\documentclass[aps,pra,showpacs,superscriptaddress,twocolumn]{revtex4}

\usepackage{graphicx,amssymb,amsmath,color}

\begin{document}
\title{Waiting time distribution for trains of quantized electron pulses}
\author{M. Albert}
\affiliation{Universit\'e de Nice Sophia-Antipolis, INLN, CNRS, 06560 Valbonne, France}
\author{P. Devillard}
\affiliation{Aix Marseille Universit\'e, CNRS, CPT, UMR 7332, 13288 Marseille, France}
\affiliation{Universit\'e de Toulon, CNRS, CPT, UMR 7332, 83957 La Garde, France}

\begin{abstract}

We consider a sequence of quantized Lorentzian pulses of non-interacting electrons impinging on a quantum point contact (QPC) and study the waiting time distribution (WTD), for any transmission and any number of pulses. As the degree of overlap between the electronic wave functions is tuned, the WTD reveals how the correlations between particles are modified. In the weak overlap regime, the WTD is made of several equidistant peaks, separated by the same period as the incoming pulses, contained in an almost exponentially decaying envelope. In the other limit, the WTD of a single quantum channel subjected to a constant voltage is recovered. In both cases, the WTD stresses the difference between the fluctuations induced by the scatterer and the ones encoded in the incoming quantum state.  A clear cross-over between these two situations is studied with numerical and analytical calculations based on scattering theory. 
\end{abstract} 
\pacs{73.23.-b, 73.63.-b, 72.70.+m.}  
\maketitle
\section{Introduction}

The past decade has been marked by the emergence of electron quantum optics. In the spirit of quantum optics with photons, it aims to generate and manipulate single electronic excitations in quantum coherent circuits for fundamental and applied science. As a first step to achieve this goal, several single electron sources have been implemented in sub-micron cavities, such as the so called quantum capacitor \cite{Gabelli06,Feve07,Buttiker93} and others \cite{Blumenthal,Pekola,Giazotto,Leight,Hermelin,McNeil,Rogge2014}, or by applying a periodic sequence of Lorentzian voltage pulses to an electronic reservoir in order to generate a clean and coherent train of electronic excitations \cite{Levitov96,Ivanov97,Keeling2006,Dubois}. Once injected into quantum circuits, such excitations can be used to study fundamental aspects of quantum mechanics such as entanglement \cite{Splett2009}, interference effects \cite{Ol2008, Juergens2011} and quantum correlations \cite{Jonck2012,Bocquillon2012,Bocquillon2013} or interaction effects \cite{Bocquillon2013_2} and coherence properties \cite{Degiovanni2009,Haack2011} which would be of great interest when it comes to applications in quantum electronics or information processing. 

However, due to quantum effects, it is now well established that charge transport at the nanoscale is a statistical process \cite{Buttiker2000}. Going beyond the knowledge of average currents is then unavoidable and extremely useful at the same time as pointed out by R. Landauer in his famous quote ``the noise is the signal''. Therefore, many efforts have been made in this direction in the past two decades using noise measurements \cite{Buttiker2000} and full counting statistics (FCS) \cite{Buttiker2000,Levitov93,Nazarov2003,Andreev2000,Makhlin2001,Albert2011}, namely the second moment of current fluctuations and the statistics of charges transferred during a long time interval. Recently, other tools have been introduced to characterize current fluctuations in the time domain like the finite-frequency noise \cite{Buttiker2000,Basset,Mahe2011,Parmentier2012,Zamoun2012} and FCS \cite{Emary2007,Flindt2008,Marcos2010,Marcos2011,Ubbelohde12}, the Wigner function \cite{Ferraro2013} or the waiting time distribution (WTD) \cite{Scriefl05,Koch2005,Welack2008,Brandes2008,Albert11,Albert12,FlindtThomas,Dasenbrook,Rajabi2013,FlindtThomas2,Wang2014}. Indeed, in such quantum devices, the time between the detection of two consecutive electrons is random because of their quantum nature and the knowledge of its probability distribution provides an original point of view on quantum correlations and current fluctuations.

\begin{figure}
  \includegraphics[width=0.9\linewidth]{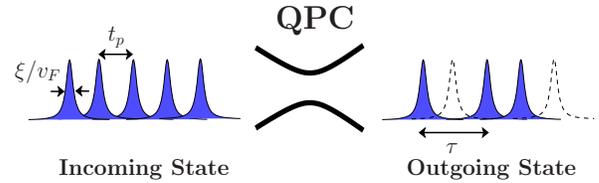}
  \caption{(Color online) Schematic picture of the system: a train of Lorentzian wave packets is emitted by a periodically driven reservoir and propagates to the right toward a QPC of transparency $T_e$. The outgoing many-body state is made of transmitted and missing (reflected) electrons represented by filled curves and dashed lines respectively. The quantity of interest is the distribution of delay times $\tau$ between the detection of two electrons far away from the scattering region.\label{fig_setup}}
\end{figure}

We study here the WTD of non-interacting electrons emitted in a sequence of Lorentzian pulses as proposed theoretically \cite{Levitov96,Ivanov97,Keeling2006} and demonstrated experimentally recently \cite{Dubois}. We focus here on the regime where electrons are emitted one by one when applying a periodic voltage to an electronic reservoir in a clean fashion (no spurious electron-hole pairs). These electrons are then propagating along a one-dimensional quantum channel and may scatter onto a quantum point contact (QPC) before being detected (see Fig. \ref{fig_setup}). This situation is physically different from the one studied in Ref. \onlinecite{Albert12} where free electrons were injected directly from the reservoir subjected to a constant bias $eV$ into a single channel. The ratio $R$ between the width of the wave packets $\xi$ and the interparticle distance $v_F t_p$ is a new control parameter that drives the many-body quantum state from a solid-like state, when the wave packets weakly overlap, to a liquid-like behavior at strong overlap. In the latter limit, the constant bias case is recovered. We use a wave packet approach  \cite{Landauer87,Hassler08} to derive general formulas for the WTD that we evaluate both numerically and analytically in some limiting cases. These formulas are equivalent to those derived in \cite{Dasenbrook} from a second-quantized formulation. In addition we show that the decay rate of the WTD corresponds to the one of a binomial process and describe analytically the quantum fluctuations around this classical result. Although we focus on Lorentzian pulses here, our theory is valid for arbitrary shapes of the wave packets.

This paper is organized as follows. In Sec. II we describe the setup under investigation and show how to compute the WTD from the many-body state and the scattering matrix of the quantum point contact. We then turn to the results obtained with this formalism, starting from the two-particle case for pedagogical reasons in Sec. III before discussing the periodic case in Sec. IV. Section V is devoted to a brief discussion of experimental measurements of the WTD. Conclusion and outlook are given in Sec. VI and several technical details are available in appendices.

\section{Model}

We consider a QPC connected to two one-dimensional electronic leads. A periodic sequence of Lorentzian pulses of the form

\begin{equation}
  V(t)=\frac{\hbar}{e}\sum_{n=-\infty}^{+\infty}\frac{2t_\xi}{(t-n\, t_p)^2+{t_\xi}^2},
\end{equation}
with $t_\xi=\xi/v_F$, is locally applied to the left reservoir and generates a train of single electron excitations without electron-hole pairs as demonstrated theoretically \cite{Keeling2006} and experimentally \cite{Dubois} recently. Here $\xi$ and $v_F t_p$ will be respectively the width and the spatial period of the wave packets. We focus here on the special case where each pulse contains exactly one electron. These electrons are moving to the right and scatter on the QPC as depicted in Fig. \ref{fig_setup}. In addition we consider zero temperature, therefore only electrons emitted from the left reservoir are involved in the transport process.
  
The incoming train of electrons, generated by the periodic voltage, is built up from a set of $N$ single particle Lorentzian wave packets separated by time interval $t_p$,
\begin{equation}
\psi_\ell(x,t) = \sqrt{\frac{\xi} {\pi}} \frac{i}{x-v_F (t-\ell t_p) + i \xi},
\label{spwf}
\end{equation}
$\ell=0,N-1$. At low energy, the dispersion relation is linear and the Fermi velocity $v_F$ is supposed to be a constant (independent of energy). Each wave packet is a superposition of plane waves of energy $E_k$ but the amplitude in the superposition is exponentially decaying with energy (or wave number since $E_k=\hbar v_F k$). A crucial quantity is the overlap between wave functions $R_{\ell,\ell^{\prime}}=\langle \psi_\ell \vert \psi_{\ell^{\prime}} \rangle$, for $\ell \not= \ell^{\prime}$ that controls the correlations between the different electrons. Taking into account the fermionic statistics, the many-body wave-function has to be antisymmetrized and as far as interactions are negligible, it will be given by a Slater determinant of all the possible orbitals $\psi_\ell$.

Now, the pulses coming from the left of the scattering regions (see Fig.\ref{fig_setup}) are impinging on a quantum point contact of transparency $T_e$, so that $\psi_\ell$ needs to be changed to

\begin{equation}
  \psi_\ell(x,t)=\int_0^{+\infty} \,\sqrt{\frac{\xi}{\pi}}\, t_k\,e^{-\xi k}\,e^{ik(x+\ell v_F t_p-v_F t)}\,dk, 
\end{equation}
with $T_e=|t_k|^2$, for $x$ far away to the right of the QPC, and will be measured by a detector located in this region. For the sake of simplicity,
 we only consider energy independent scattering with therefore $t_k=\sqrt{T_e}$.

For negligible overlap, the normalization is trivial apart from an overall $\sqrt{N!}$ factor. However, for finite overlap, this statistical coefficient has to be completed with the determinant of the overlap matrix elements between different wave packets \cite{Hassler08} $ \, \sqrt{\textrm{det}(R_{\ell,\ell^{\prime}})}$. 

 As mentioned in the introduction, the quantity of interest is the WTD $\mathcal W(\tau,t_0)$, namely the probability distribution of delay times between the detection of two consecutive electrons. The detector is located at $x=x_0$ somewhere far away from the scattering region between the QPC and the right reservoir. In that case, the WTD depends on both the delay time $\tau$ and a second time $t_0$ which is chosen here to be the time when the first electron is measured. This situation is more general than the one of a stationary flow of particles described in Ref. \onlinecite{Albert12} where it only depends on the delay time. To calculate this quantity, we will either refer to the joint probability of measuring an electron at time $t_0$ and nothing until $t_0 +\tau$, $P(\tau,t_0)$ or the probability of not detecting anything between $t_0$ and $t_0+\tau$, $\Pi(\tau,t_0)$ even if they are related quantities. Following these definitions, it is straightforward to show the following useful relations

\begin{equation}\label{eq_wtd_P}
  \mathcal{W}(\tau,t_0) \, =\, - \frac{\partial P(\tau,t_0)} {\partial \tau},
\end{equation}
\begin{equation}
p(t_0)\,P(\tau,t_0) \, =\, 
\frac{\partial \Pi(\tau,t_0)} {\partial t_0}
 \, - \, \frac{\partial \Pi(\tau,t_0)}{\partial \tau},
\label{wtaumethod2}
\end{equation}   
and
\begin{equation}
p(t_0)\,{\cal W}(\tau,t_0) \, =\, 
\frac{\partial^2 \Pi(\tau,t_0)}{\partial \tau^2} \, - \, 
\frac{\partial^2 \Pi(\tau,t_0) }{\partial \tau \partial t_0},
\label{wtaumethod2_2}
\end{equation}   
where $p(t_0)$ denotes the probability density to detect an electron at time $t_0$ and is simply proportional to the average current. Although not fundamentally different, we will use either $P$ or $\Pi$ to compute the WTD depending on mathematical convenience. In a real experiment, the time of the first detection is random. However, for periodic systems ($N\gg 1$), one can construct an average WTD that only depends on the time delay $\tau$. Such a quantity is constructed from the time integration of the WTD over a period with weight $p(t_0)$. Using this definition and (\ref{wtaumethod2_2}) yields

\begin{equation}
  \mathcal{\overline{\mathcal{W}}(\tau)}=\langle \tau \rangle\frac{d^2\,\overline{\Pi}(\tau)}{d\tau^2}\, ,
\end{equation}
with $\overline{\mathcal{W}}(\tau)=\int_0^{t_p} p(t_0) \mathcal W(\tau,t_0) dt_0 / \int_0^{t_p} p(t_0) dt_0$, $\overline{\Pi}(\tau)=\int_0^{t_p} \Pi(\tau,t_0) dt_0/t_p$ and $1/\langle\tau\rangle=\int_0^{t_p} p(t_0) dt_0/t_p$ the mean waiting time. This is the exact analog of the formula for a stationary process \cite{Albert12} and the same formula as the one proposed by Dasenbrook \textit{et al.} \cite{Dasenbrook} in a recent related work.

For pedagogical reasons, we will start to explain the calculations in the case $N=2$ before describing the physics of a periodic state made of $N\gg 1$ electrons.

\subsection{Two-electron case}

The normalized wave-function for the two electrons is simply 
\begin{equation}
\psi_S \, =\, \frac{1}{\sqrt{2}} \frac{ 1}{\sqrt{ D_r}}
 \lbrack \psi_1(x_1) \psi_2(x_2) - \psi_2(x_1) \psi_1(x_2) \rbrack ,
\end{equation}
with 

\begin{equation}\begin{split}
{\cal D}_r \, = & \, ( \vert \psi_1(x_0) \vert^2 + \vert \psi_2(x_0)\vert^2 \\
& - 2 \textrm{Re}[\psi_1^*(x_0) \psi_2(x_0) \langle \psi_2 \vert \psi_1 \rangle])^{1/2}.
\end{split}
\end{equation}   

The detector is located at $x_0$, in a very small interval. Suppose we measure an electron at $x_0$, at time $t_0$. The new wave function is obtained by acting the operator $Q_1 = \int_{x_0 - v_F t_u}^{x_0} \vert x \rangle \langle x \vert \, dx$, where $t_u$ is a very small time, much smaller than both $t_p$, the interval between the pulses and $\xi/v_F$. 

After the measurement, the packet is reduced, so that the wave-function of one electron which has been detected is now $\varphi(x)$, where $\varphi(x)$ is very peaked around the detector and almost zero everywhere else. We can take $x=x_1$, so that $\int_{- \infty}^{\infty} \vert\varphi(x_1-x_0) \vert^2 \, dx_1=1$. After a few algebraic manipulations, the many-body wave function after the measurement $\vert \Psi_{Sa} \rangle$ assumes the form

\begin{equation}\begin{split}
\vert \Psi_{Sa} \rangle \, = & \, 
\frac{1}{{\cal D}_r} \vert \varphi(x_1-x_0) \rangle\\
& \otimes \bigl\lbrack \psi_1(x_0) \vert \psi_2(x_2)\rangle -
 \psi_2(x_0) \vert \psi_1(x_2) \rangle \bigr\rbrack .
\end{split}\end{equation}

Therefore, the probability of detecting nothing before time $\tau$, having detected an electron at time $t_0$, is 
\begin{equation}
P(\tau, t_0) \, =\, \Bigl \langle  \Psi_{Sa} 
\biggl\vert
\Bigl( 1 - \int_{x_0 - v_F \tau}^{x_0}
 \vert x^{\prime}_2 \rangle \langle x^{\prime}_2 \vert \, dx^{\prime}_2 \Bigr)
\biggr\vert
\Psi_{Sa} \Bigr\rangle,  
\end{equation}       
and the WTD, through (\ref{eq_wtd_P}), reads
\begin{equation}
\mathcal{W}(\tau , t_0) = \frac{v_F}{ {\cal D}_r^2} 
\bigl\vert \psi_1(x_0) \psi_2(x_0 - v_F \tau) - 
 \psi_2(x_0)\psi_1(x_0 - v_F \tau) \bigr\vert^2.
\label{wtd2e1}
\end{equation} 

As a consequence, if the single particle wave functions $\psi_\ell$ are differentiable, then $\mathcal{W}(\tau,t_0)$ vanishes as $\tau^2$ for small waiting times, in accordance with the Pauli principle. A more expanded discussion of this result is given in Sec. III.         
 
\subsection{$N$-electron case}

In order to mimic an infinite train of Lorentzian pulses, we generalize the previous calculation to an arbitrary number of electrons $N$. We then analyze the asymptotic properties of the WTD for large $N$. Two different physical situations will be treated separately for mathematical convenience. In Sec. II B 1, we consider the situation where the single particle wave functions weakly overlap in real space. Such a situation is more easily tackled in the basis of wave-functions $\psi_\ell$ in real space. However, for large overlap, the matrices that appear in the calculation have very small determinants and are thus very ill-conditioned
 for numerical calculations. Therefore, it is more convenient mathematically to use another basis. Such a basis is constructed from the Fourier transforms of the original localized wave packets \cite{Hassler08} as explained in Sec. II B 2.

\subsubsection{Real space}

We derive here a general formula for the WTD of a train of $N$ electrons in terms of the ratio of different determinants. This formula is general but not convenient in the limit of large overlap between the single particle wave functions, as we will discuss later. The first step of the derivation consists in computing the many-body wave function after the measurement of the first particle.

Before the detection, the system is in state $\Psi_b \, =\, \frac{1}{\sqrt{N!}}\, \textrm{det} M_{i,j}/\sqrt{\textrm{det} R_{i,j}}$, with $M_{i,k}$ the matrix of elements $\psi_i(x_j)$ and $R_{i,j}$ is the overlap matrix $\langle \psi_i \vert \psi_j \rangle$. Immediately after the measurement, the state collapses through the application of a projector onto the state $\vert \varphi(x_0 -x_1) \rangle$, i.e. one electron is now confined in the detector. After normalization and a little algebra, the new wave function can be cast as
\begin{equation}
\Psi_a \, =\, \textrm{det}({\cal M}_{i,j})/\sqrt{\textrm{det} {\cal R}_{i,j}},
\end{equation}
where ${\cal M}_{i,j}$ is the same matrix as $M_{i,j}$ but the first line $\psi_1(x_1), \psi_2(x_2),..., \psi_N(x_N)$ has been replaced by $\psi_1(x_0) \varphi(x_1-x_0) , \psi_2(x_0) \varphi(x_2-x_0),...,\psi_N(x_0) \varphi(x_N-x_0)$. ${\cal R}_{i,j}$ is the same matrix as $R_{i,j}$ except that the first line has been replaced by $\langle \psi_i \vert Q(t_u) \vert \psi_j \rangle$, $j=1$ to $N$. $Q(t_u)$ is the operator $\int_{x_0 - v_F t_u}^{x_0} \vert x \rangle \langle x \vert \, dx$, namely the projector on the detector.

The probability of not detecting anything before $\tau$ is thus
\begin{equation}\begin{split}
P(\tau,t_0) \, = &\, 
\biggl\langle
 \Psi_a 
\biggl\vert 
\Bigl(1 - \int_{x_o - v_F \tau}^{x_0} \vert x \rangle \langle x \vert \, dx \Bigr)_1 \otimes
... \\
& \otimes
\Bigl(1 - \int_{x_o - v_F \tau}^{x_0} \vert x \rangle \langle x \vert \, dx \Bigr)_N
\biggr\vert
\Psi_a
\biggr\rangle,
\end{split}
\end{equation}
where the subscript $i$ means operation on coordinate $x_i$. Detailed calculations can be found in appendix A. Expanding the wave function $\Psi_a$, which is a determinant, as a sum over permutations, we arrive at
\begin{equation}
P(\tau,t_0) \, =\, 
{\sum_{\ell=1}^N \textrm{det}({\cal N}_\ell})/
{\sum_{\ell=1}^N \textrm{det}({\cal D}_\ell}),
\label{generalformula}
\end{equation}  
where ${\cal N}_\ell$ and ${\cal D}_\ell$ are two matrices defined as follows. ${\cal N}_0$ is the matrix $\langle \psi_i \vert 1 - Q \vert \psi_j \rangle$. ${\cal N}_\ell$ is obtained from ${\cal N}_0$ by substitution of the $\ell^{th}$ line by the line vector $ V_\ell \, =\, \langle \psi_\ell \vert Q(t_u) \vert \psi_1\rangle , ...,\langle \psi_\ell \vert Q(t_u) \vert \psi_N\rangle$. ${\cal D}_0$ is the overlap matrix $\langle \psi_i \vert \psi_j\rangle$ and ${\cal D}_\ell$ is again obtained from ${\cal D}_0$  by substituting the $\ell^{{\rm th}}$line by $V_\ell$. Equation \ref{generalformula} is the central result of this section and is equivalent to the determinant formula derived in \cite{Dasenbrook} from a second-quantized formulation.

\subsubsection{Fourier space}

For large overlaps, in order to perform efficient and reliable numerical simulations, we need another line of attack. This is the reason why we adapt in this case the more traditional methods used previously for QPC \cite{Hassler08,FlindtThomas,Albert12}. It has also the benefit of providing other methods for analytical calculations, as explained in the following. 

The central quantity is the idle time probability $\Pi(\tau,t_0)$. It is the probability of not detecting anything in the time interval $\lbrack t_0 \, , t_0 + \tau \rbrack$, irrespective of what happens at time $t_0$. We stress that
$\Pi(\tau,t_0)$ is a different quantity from $P(\tau,t_0)$, as explained before. 

When the $N$ electrons are in states $\psi_1,...,\psi_N$, the formula giving $\Pi(\tau,t_0)$ reads
\begin{equation}
\Pi(\tau,t_0) \, = \frac{\textrm{det}(R_{i,j}- T_e Q_{i,j})}  {\textrm{det}(R_{i,j})},
\label{formulapitau2}
\end{equation}
where $Q_{i,j}$ is $\langle \psi_i \vert Q \vert \psi_{j}\rangle$. However, as it stands, in real space basis, $\textrm{det}\,(R_{i,j})$ is very small and leads to numerical problems for large overlaps $R$. 

We thus switch to Fourier representation and define 
\begin{equation}
\phi_K \, =\, \frac{1}{\sqrt{N}} \sum_{\ell=0}^{N-1} 
\exp\bigl(-i \ell\, \frac{2 \pi}{N} K\bigr) \, \psi_\ell\,,
\end{equation}
with $K\in [0,N-1]$. For large $N$ and for $x$ in the bulk of the train, the $\phi_K$'s assume the form
\begin{equation}
\begin{split}
\phi_K \,= \, 
 \frac{1}{v_F t_p} \sqrt{\frac{\xi}{\pi}} \, \sum_{l=- \infty}^{\infty}
\Theta\Bigl(\frac{K}{N} + l \Bigr)\, 
e^{- \frac{2 \pi \xi}{v_F t_p} l} \\
\exp \left\{- 2 i \pi \Bigl(\frac{K}{N} + l \Bigr) 
\Bigl(\frac{x}{v_F t_p} - \frac{t-t_0}{t_p}\Bigr)
\right\} \, e^{- \frac{2 \pi \xi}{v_F t_p} \, \frac{K}{N}}\, ,
\end{split}
\label{formulaphiK}
\end{equation}
where $\Theta$ is the Heaviside function. Since the $\phi_K$'s are delocalized, within the span $N t_p$ of the whole train, supposed to be the largest distance (much larger than $\xi$), the $\phi_K$'s are not normalized. Thus we defined normalized functions  ${\tilde \phi}_K$'s by ${\tilde \phi}_K \, \equiv \, \phi_K/\sqrt{\langle \phi_K \vert \phi_K \rangle}$. The ${\tilde \phi}_k$'s have essentially the same form as the $\phi_\ell$'s, see Eq. (\ref{formulaphiK}), except that the real exponential factor $\exp(- 2 \pi \frac{\xi}{v_F t_p} \frac{K}{N})$ is now absent. In full analogy to Eq. (\ref{formulapitau2}), we have the formula
\begin{equation}\label{eq_pi_fourier}
\Pi(\tau,t_0) \, = \, \frac{\textrm{det}({\cal R}_{K,K^{\prime}} - T_e {\cal Q}_{K,K^{\prime}})}{\textrm{det}({\cal R}_{K,K^{\prime}})},
\end{equation}
where the matrices ${\cal R}_{K,K^{\prime}}$ and ${\cal Q}_{K,K^{\prime}}$ are the exact analogs of $R_{K,K^{\prime}}$ and $Q_{K,K^{\prime}}$ previously defined; the $\psi_k$'s just need to be replaced  by the  ${\tilde \phi}_K$'s.  In the limit of large $N$ (only), the elements of the matrices ${\cal R}_{K,K^{\prime}}$ and ${\cal Q}_{K,K^{\prime}}$ depend solely on $K-K^{\prime}$, they are Toeplitz matrices. This leads to great simplifications, both analytically and numerically.

\section{Results for two pulses}

Although mathematically trivial, it is instructive to discuss the behaviors of $\mathcal W(\tau,t_0)$ for two pulses in limiting cases. We focus here on the results at perfect transmission, probing only the fluctuations encoded in the many-body state. From now, we assume that the first electron has been detected at $x=0$  at time $t=0$. The position of the center of the first and the second wave-packet are $x_1$ and $x_2 = x_1-v_F t_p$ respectively. Using (\ref{spwf}) and (\ref{wtd2e1}) yields the following compact formula
\begin{equation}\begin{split}
    \mathcal{W}(\tau,t_0) \, =& \,     2 \, \frac{v_F}{\xi}\, 
\left| \frac{1}{(X_0 +i)(X_0-R^{-1}-Y+i)}\right. \\
& \left. -\frac{1}{(X_0 -R^{-1} +i)(X_0-Y+i)} \right|^2\, D^{-1}\,, 
\end{split}
\end{equation}
where $X_i=x_i/\xi$ ($i=0,1,2$), $Y=v_F \tau/\xi$, $R=\xi/v_F t_p$ and

\begin{equation}
  \begin{split}
  D \,= & \, \frac{1}{1+X_1^2} +\frac{1}{1+X_2^2}\\
  & - 2 \textrm{Re}\biggl\lbrace \frac{1}{(X_1-i)(X_2-R^{-1}+i)(1+i R^{-1})} \biggr\rbrace.
  \end{split}
\end{equation}

These formulas are not particularly illuminating, thus, we consider two limiting cases. 

In the weak overlap case, $R \ll 1$, and for small times we get 
\begin{equation}
\mathcal{W}(\tau,t_0) \simeq \,  \frac{2 v_F}{\xi} \, \frac{1}{1 + (\frac{x_0 - v_F \tau}{\xi})^2} \, 
\Bigl(\frac{\tau}{t_p}\Bigr)^2,
\end{equation}
which indeed vanishes as $\tau^2$, as expected from the Pauli principle. In contrast, for large times, we find
\begin{equation}
  \mathcal{W}(\tau,t_0) \simeq 2 \,\frac{\xi}{v_F} \, \tau^{-2},
\end{equation}
which decays algebraically with $\tau$. This is a direct consequence of the shape of the wave function of the emitted electrons. Since there are only two electrons, the last one is not correlated with any other following electron and keeps its Lorentzian tail. Physically, it means that the fluctuations of the waiting time can be arbitrary large and are not characterized by a finite second moment. Such a behavior is not possible in the bulk of a train of $N\gg 1$ electrons. Indeed, all the electrons have to be contained in a time window of order $N t_p$ which prevents large waiting times fluctuations. In addition we find that the maximum of $W(\tau,t_0)$ occurs for a time of the order $t_p$. 

In the opposite case of strongly overlapping pulses, $R \gg 1$, we get
\begin{equation}
\mathcal{W}(\tau,t_0) \simeq \frac{2}{3} \frac{v_F}{\xi} \, 
\Bigl\lbrack\frac{1}{1 + (\frac{x_0 - v_F \tau}{ \xi})^2} \Bigr\rbrack^2
\Bigl(\frac{\tau}{ \xi/v_F}\Bigr)^2.
\end{equation}
Again, $\mathcal{W}(\tau,t_0)$ starts as $\tau^2$ for small times and decays as $\tau^{-2}$ at large times. However, now, the only relevant time is $\xi/v_F$, which is related to the span of wave-packets. Time $t_p$ has completely scaled out of the problem.

\section{Results for large $N$}

We now discuss the results obtained for an infinite train of electrons by looking at the large $N$ limit. This section starts with the results for weakly overlapping wave packets where the WTD presents a very clear internal structure made from periodic peaks every $t_p$. As the overlap is increased, such strong solid-like correlations smoothly disappear until becoming small wiggles as expected for free fermions propagating in a single quantum channel \cite{Albert12}. We recall that the overlap parameter is defined as $R=\xi/ (v_F t_p)$.

\subsection{Weakly overlapping wave packets}

\begin{figure}
  \includegraphics[width=0.9\linewidth]{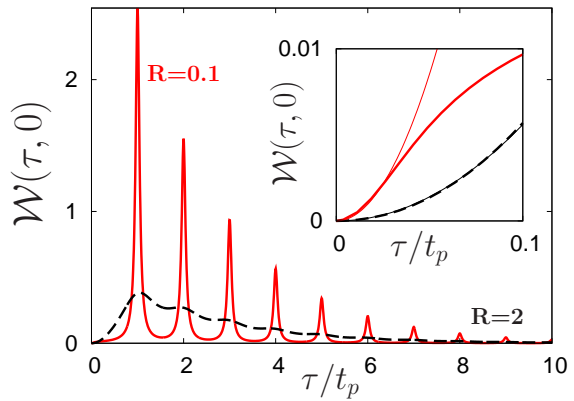}
  \caption{(Color online) WTD versus $\tau/t_p$ for $R=0.1$ (solid red line), and $R=2$ (dashed-dotted black line). $T_e=0.4$ and the number of pulses is always $N=50$. Inset: enlargement of the small time region compared to quadratic fitted functions (thin solid lines) to show the $\tau^2$ dependence.\label{fig_wtd1}}
\end{figure}

In this regime, the train of particles resembles a series of well separated thin Lorentzian wave packets. These wave packets are almost independent and periodically distributed in time. We therefore expect to observe well defined peaks in the WTD, reflecting the shape of a single particle wave function. At perfect transmission, there must be (approximately) only one peak centered around $t_p$. However, for finite transmission the WTD exhibits a double structure made of an envelope function containing several similar peaks. Roughly speaking, these peaks are the hallmark of the fluctuations encoded in the quantum state whereas the envelope is coming from the random scattering through the quantum point contact. Such a situation is similar to what has been observed for the WTD distribution of a single electron source in the phase noise regime \cite{Albert11}. 

We perform numerical calculations using Eq. (\ref{generalformula}) which is more convenient in this regime. Figure \ref{fig_wtd1} shows $\mathcal{W}(\tau,0)$ for two values of $R$ and confirms our expectations. Moreover, the small time behavior is shown to correspond to the $\tau^2$ prediction imposed by Pauli principle. Such a behavior is indeed observed for times shorter than $R t_p$. Note that we show $\mathcal{W}(\tau,t_0)$ for $t_0=0$ but the results are qualitatively similar for any value of this parameter.

We now show how to obtain an approximant of the WTD in the limit of extremely small overlap and long waiting times. In that case, Eq. (\ref{generalformula}) reduces for large times to 
\begin{equation}
P(\tau,t_0) \, = \,  \frac{\textrm{det}(R_{\ell,\ell^{\prime}} - T_e N_{\ell,\ell^{\prime}})}{\textrm{det}(R_{\ell,\ell^{\prime}})},
\end{equation}
with
\begin{equation}
  R_{\ell,\ell^{\prime}} \, =\, \frac{1}{1 - i \frac{(\ell^{\prime}-\ell)}{2 \pi R}}, 
\end{equation}
\begin{equation}
  \begin{split}
    N_{\ell,\ell^{\prime}} \,=& \, \frac{R_{\ell,\ell^{\prime}}}{2 \pi}\, 
    \Bigl\lbrack
    \tan^{-1} \bigl( \ell^{\prime} R^{-1}\bigr) - 
    \tan^{-1} \bigl(\lbrack \ell^{\prime}-(\tau/t_p)\rbrack R^{-1}\bigr) \\
    &+ (\ell \leftrightarrow \ell^{\prime})\,
    -\frac{i}{2} \ln(f_\ell/f_{\ell^{\prime}})
    \Bigr\rbrack,
  \end{split}
\end{equation}
 and $f_\ell =(\ell^2+ R^2)/\lbrack(\ell - \tau/t_p)^2 +R^2\rbrack$. For vanishing overlap we throw out the non diagonal elements because only $R_{\ell,\ell}$ is non negligible. Such a naive computation gives (see appendix B for details)
\begin{equation}
  \begin{split}
    \mathcal{W}(\tau,t_0) \simeq \sum_n\delta(\tau+t_0-nt_p)  \exp\Bigl(\ln(1 - T_e) \, \textrm{int}(\frac{\tau+t_0}{t_p})\Bigr),
  \end{split}
  \label{simpleapproximation}
\end{equation} 
where ``$\textrm{int}$'' denotes the integer part. This result is physically illuminating and allows us to extract the decay rate of the WTD. At this level of approximation, the WTD reduces to a periodic series of peaks contained in an exponentially decaying envelope. The presence of several peaks is due to the imperfect transmission through the QPC that allows electrons to be reflected. Actually, this is the WTD of a classical binomial process. The only random process is the rate of success $T_e$ every $t_p$ and the WTD provides only information about the scatterer. Quantum corrections would give a finite width to the peaks as a hallmark of quantum jittering. Such information are encoded in the many-body state and in that case would lead to a Lorentzian shape even though we have not proved it explicitly. However, when $R\to 0$ we can assume that the electrons are uncorrelated and the shape of the peaks of the WTD just reflects the one of the wave packets. This yields

\begin{equation}\label{wtd_ap1}
   \mathcal{W}(\tau,t_0) \simeq \sum_n w_n(\tau+t_0)  \exp\Bigl(\ln(1 - T_e) \, \textrm{int}(\frac{\tau+t_0}{t_p})\Bigr),
\end{equation}
with
\begin{equation}\label{wtd_ap2}
  w_n(\tau)=\frac{R}{\pi}\frac{t_p}{(\tau-n\,t_p)^2+t_p R^2}\,,
\end{equation}
which is in very good agreement with numerical calculations for $R\ll 1$ (see Fig. \ref{fig_trans} for $T_e=1$ and $R=0.02$). The decay rate extracted numerically for $T_e=0.4$ and $R=0.1$ is $0.5106$, in very good agreement with the prediction $-\ln(1-T_e)=0.5108$. Figure \ref{fig_decay} presents a more detailed comparison for different transmissions. However, as $R$ becomes larger, the peaks continue to spread and eventually overlap. As explained in the next subsection, this modifies the correlations between particles and breaks the uncorrelated electrons picture.

At perfect transmission, the crude approximation leading to Eq. (\ref{simpleapproximation}) is not valid anymore. Since there is no scattering, a particle is almost surely found every $t_p$. Therefore, the WTD consists in a single peak centered around $t_p$ plus a very small satellite around $2t_p$ as depicted on Fig. \ref{fig_trans}. Other satellites exist at $n t_p$, with $n$ integer, but have negligible amplitudes. Moreover, Fig. \ref{fig_trans} illustrates the agreement with (\ref{wtd_ap2}) for $T_e=1$ and $R=0.02$. However, this agreement is only valid for times around $t_p$ and the asymptotic decay of the WTD is not Lorentzian but rather Gaussian.

\subsection{Intermediate and large overlaps}

We now turn to the discussion of intermediate and large values of the overlap parameter $R=\xi/v_F t_p$. We start with numerical results before giving analytical arguments in the infinite $R$ and in the 
finite but large $R$ limits. We then try to explain how the finiteness of $\xi$ alters the results.

\subsubsection{Numerical results}

\begin{figure}
\includegraphics[width=0.9\linewidth]{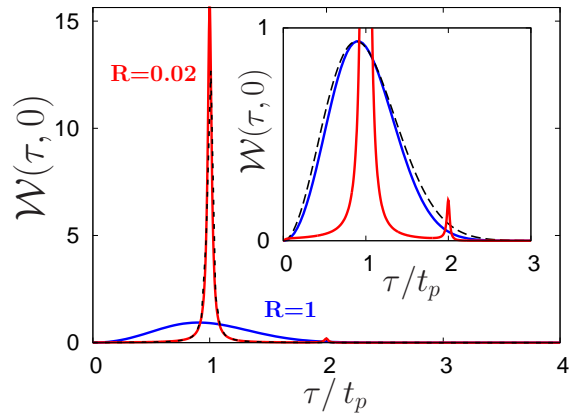}
\caption{(Color online) WTD versus $\tau/t_p$ with $t_0=0$ for $R=0.02$ (red solid line), and $R=1$ (blue solid line). $T_e=1$ and the number of pulses is always $N=100$. Black dashed lines represent (\ref{wtd_ap2}). Inset: enlargement to illustrate the agreement with the Wigner-Dyson distribution (\ref{eq_wd}), (dashed line) for $R=1$ and the presence of a second peak in the WTD for $R=0.02$.\label{fig_trans}}
\end{figure}

In this case, it is convenient to switch to the formulation in Fourier space described in the previous section. We then use (\ref{eq_pi_fourier}) and (\ref{wtaumethod2_2}) to compute the WTD numerically. Figure \ref{fig_wtd1} shows the result for $R=2$ and $T_e=0.4$. We clearly see that the correlations are strongly reduced since the peaks melt down to small wiggles. These small oscillations are still the hallmark of periodic correlations encoded in the periodic train of electrons. The situation is comparable to a liquid where correlations on the scale of the inter-particle distance are much weaker and decay rapidly. Then, as $R$ is tuned to large values, we recover the case \cite{Albert12} of a single quantum channel subjected to an effective constant bias $eV=h/t_p$ as pointed 
out in Ref. \onlinecite{Hassler08}. Indeed, we already see for $R=1$ and $T_e=1$ on Fig. \ref{fig_trans} that the finite overlap case approaches the infinite case. However, a closer look at the asymptotic properties shows that for times of the order or larger than $R t_p$ the decay rate, although still Gaussian, is different.

To get a better understanding of this, we study analytically the decay rate and the wiggles for infinite and finite $R$ using the theory of Toeplitz matrices.

\subsubsection{Infinite overlap}

As mentioned above, the limit of infinite $\xi$ is equivalent to a biased single quantum channel filled with free fermions \cite{Hassler08}. Numerical calculations, as well as a few limiting cases such as QPC close to pinch-off, have been presented in Ref. \onlinecite{Albert12}. However, using Szeg\"o's theorem, we can evaluate the envelope and the wiggles in the long time limit with good accuracy for $T_e\ne 1$. The case of perfect transmission has to be tackled aside.

We recall that, in this limit, $\mathcal Q_{K,K'}$ that appears in the determinant formula (\ref{eq_pi_fourier}) simplifies to

\begin{equation}
{\cal Q}_{K,K^{\prime}} \simeq e^{i \frac{2 \pi}{N} (K-K^{\prime}) \frac{x_0}{v_F t_p}} \, 
\frac{\exp(2 i \pi \frac{K-K^{\prime}}{N} \frac{\tau}{t_p})\, - 1}{2 i \pi (K-K^{\prime})}.
\end{equation} 
Moreover, in real space $\langle \psi_\ell \vert \psi_{\ell^{\prime}}\rangle$ is a Lorentzian function decaying on scale $\xi$, therefore in Fourier space, the magnitude of the elements of ${\cal R}_{K,K^{\prime}}$ decays exponentially on scale $N/R$. For infinite overlap, i.e. $\xi \gg v_F t_p$ but still $\xi \ll N t_p$, ${\cal R}_{K,K^{\prime}}$ goes to the identity matrix. It is convenient to introduce the symbols $r(\theta)$ of the matrices ${\cal R}_{K,K^{\prime}}$ defined as
\begin{equation}
r(\theta) \, =\, \sum_{l=-N}^N e^{i\, l \theta} {\cal R}_{K,K^{\prime}},
\end{equation} 
with $l=K-K^{\prime}$ and $\theta$ taken in $\lbrack - \pi , \pi\rbrack$ and analogously, $q(\theta)$ the symbol of ${\cal Q}_{K,K^{\prime}}$. For infinite overlap and setting $x_0=0$ \cite{notex0}, we find $q(\theta) = 1$ if $\vert \theta \vert < \pi \frac{X}{N} $ and $0$ otherwise, with $X=\tau/t_p$ and $r(\theta)$ is $1$ for any $\theta$.  

\begin{figure}
\includegraphics[width=0.9\linewidth]{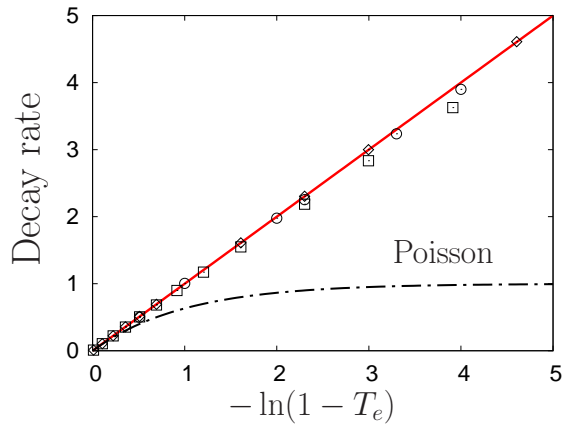}
\caption{(Color online) Decay rate of the WTD for $R=0.1$ ($\diamond$), $R=2$ ($\square$) and $R=+\infty$ ($\circ$) as a function of $-\ln(1-T_e)$ in agreement with (\ref{wtd_ap1}) and (\ref{wtd_ap3}). Dashed-dotted black line is the Poissonian result which states that for small transparency the decay rate is simply $T_e/t_p$.\label{fig_decay}}
\end{figure}

To obtain the long time behavior of $\Pi(\tau,t_0)$ using Eq. (\ref{eq_pi_fourier}), we use Szeg\"o's theorem which gives the long time asymptotic behavior of Toeplitz determinants. This reads as
\begin{equation}\label{pi_ap}
\Pi(\tau,t_0)\propto \exp\left\{\, N \displaystyle{\int_{-\pi}^{\pi}} \ln [ 1 - T_e\, q(\theta)] \, \frac{d \theta}{2 \pi}\right\},
\end{equation}
which yields the main behavior of the WTD for large times.
\begin{equation}\label{wtd_ap3}
  \mathcal W(\tau,t_0) \propto \exp\Bigl(\ln(1- T_e)\, \tau/t_p\Bigr). 
\end{equation}
Again, we find that for $T_e\neq 1$, the WTD decays exponentially with a rate $\ln(1-T_e)$ and a comparison to numerical results on Fig. \ref{fig_decay} shows a very good agreement with this prediction. This is related to the fact that the physics of this system, on a long time scale, is a binomial process. Note that, in this regime, $t_0$ scales out of the problem since the system becomes stationary. 

Whenever applicable, Szeg\"o's theorem not only gives the main exponential behavior but also the next correction. Let $F_k$ be the Fourier transform of $\ln f(\theta)$, at wave vector $k=\frac{2\pi}{N} n$, for $n$ integer between $-\frac{N}{2}$ and $\frac{N}{2}$. The correction is of the form $\exp\Bigl(\sum_k k \, \vert F_k\vert^2\Bigr)$, whenever the sum converges. However, in our case, Szeg\"o's theorem is not directly applicable, because the sum on $k$ diverges logarithmically. A way around it would be to apply a theorem for matrices having Fisher-Hartwig singularities \cite{FisherHartwig,Abanov} but it yields only results at very large times, in terms of a power law correction to the exponential decay of the envelope, missing the wiggles at intermediate times. We thus apply an ultraviolet cutoff on $k$, called $k_c$, which is of order $N$. If we choose $k_c = N \gamma$ with $\gamma $ a numerical constant of order one, we obtain
\begin{equation}
\begin{split}
\Pi(\tau,0) \, \propto \, 
\frac{T_e}{\ln \vert 1 - T_e \vert} \, \exp\Bigl(X \, \ln(1-T_e) \Bigr) 
\\
 \, \exp\biggl(\frac{\ln^2(1-T_e)}{\pi} \, F(X)\biggr)
\end{split}
\end{equation}
with 
\begin{equation}
F(X) \, =\, \sum_{0<k<\gamma N} \frac{1 - \cos(2 \pi X k/N)}{2 \pi k},
\end{equation}
and $X= \tau/t_p$. In the limit of very large $N$, $F(X)$ can be expressed in terms of the cosine integral and $F(X)$ reduces to $\int_0^X \frac{1 - \cos(2 \pi \gamma z)}{2 \pi z} \, dz$. $\Pi(\tau,0)$ is always a strictly decreasing function of $\tau$ but its second derivative, $W(\tau,0)$ exhibits, for moderate $\tau/t_p$, weak wiggles with period $t_p$, due to the cosine integral. Although not shown, this prediction is in good agreement with numerical results in the long time limit.

For $T_e=1$, (but $\xi$ still infinite) the behavior is known exactly \cite{Krasovsky,TracyWidom,Albert12} from an analogy with random matrix theory. The decay at large times is mainly Gaussian with power law and higher order corrections but is well approximated by the Wigner distribution
 
 \begin{equation}\label{eq_wd}
   \mathcal{W}(\tau) \, \simeq \,  \frac{32}{\pi^2} X^2 \exp\bigl(- \frac{4}{\pi} X^2\bigr).
\end{equation} 

\subsubsection{Finite but large overlap}

For the sake of simplicity, we only consider the main term in Szeg\"o's theorem and look at the decay of the WTD at large times. Following the same method as above, Eq. (\ref{eq_pi_fourier}) translates into a formula similar to Eq. (\ref{pi_ap}), but we find more convenient to write it as a Riemann sum 
\begin{equation}
\Pi(\tau,t_0) \propto 
 \exp\Bigl( \sum_{n=-N/2}^{N/2} 
 \ln \bigl\lbrack 1 - T_e\, q (\theta_n)/r(\theta_n)\bigr\rbrack\Bigr),
\label{riemann}
\end{equation}
with $\theta_n = n \pi/N$, taking discrete values. It is thus necessary to see how the finiteness of $\xi$ alters $q(\theta)$ and $r(\theta)$. 
 
Let us begin with $r(\theta)$. The overlap between ${\tilde \phi}_K$ and ${\tilde \phi}_K^{\prime}$ is still $1$ (by definition) for $K=K^{\prime}$ but will no longer be zero for $K \not= K^{\prime}$. However, its magnitude falls to order $1/N$ as soon as $K-K^{\prime}$  is $1$ or larger. It is mainly of order $1/N$ up to $K-K^{\prime}$ of the order $N/R$.  For large values of $K-K^{\prime}$, it is still smaller and we shall neglect it. Thus, $r(\theta_n)$ will no longer be $1$ for all $\theta$'s but will be slightly modified for $\vert \theta_n \vert$ smaller than $R/N$. The change is of order $1/N$.

We now turn to $q(\theta)$. For $K-K^{\prime}$ smaller than $N/R$, ${\tilde Q}_{K,K^{\prime}}$ will assume the same form as for $R$ infinite. The reason is that, small $K-K^{\prime}$ mean large distance and the finiteness of $\xi$ just changes the short distance cutoff in real space. In Fourier space, small $K$'s do not see modifications on small scales. For $\theta$'s larger than $R/N$, $q(\theta_n)$ keeps its infinite $\xi$ form, which is $1$ for $\vert \theta \vert <\pi X/N$ and zero otherwise. Only $\theta$'s of absolute value smaller than $R/N$ deserve much more care. 

 It is natural to single out the $\theta_n$'s for which $q(\theta_n)$ and $r(\theta_n)$ retain their infinite $R$ values. We call theses values $q_{\infty}(\theta)$ and $r_{\infty}(\theta)$. As $r_{\infty}(\theta)$ is always one, we omit it. Times much larger than $R t_p$ correspond to $X \gg R$. In this case, we split the sum in Eq. (\ref{riemann}) into two parts, 
\begin{equation}
\begin{split}
\ln\, \Pi(\tau,t_0) \, \propto \, & 
\sum_{n=-R/2}^{R/2} \ln\bigl(1 - T_e\, q(\theta_n)/r(\theta_n)\bigr) \\
 & + \sum_{\vert n \vert =R/2}^{\textrm{int}(X/N)}  \ln(1 - T_e\, q_{\infty}(\theta_n))
\end{split}
\end{equation}     
  For $X \gg R$, the second sum will make the main contribution and the first one is much smaller. Thus, we expect that the rate of decay for large times will be approximately the same as the one observed for infinite overlap $R$. This is indeed what is observed numerically. Already for $R=1$, the rate of decay is very similar to the one observed for infinite $R$.  

  This argument does not work for $T_e=1$ because Szeg\"o's theorem is not applicable. Numerically, however, we always observe Gaussian behavior for perfect transmission at times larger than $R t_p$, whether $R$ is finite or not.  

  For times, smaller than $R t_p$, only $\theta's$ smaller than $R$ are involved and both $q$ and $r$ do get perturbed by the finiteness of $\xi$. We did not attempt to make analytical predictions in this time range, $1 \ll  X  \ll R$.         
\section{Experimental considerations}

We now address the question of measurability of WTD. Detecting single electron events in a quantum coherent conductors is one of the most challenging goal of electron quantum optics nowadays. Most of the single electron sources operate at the GHz range (namely they inject coherent electrons at a GHz frequency) while standard electronic technology limits the measurement precision to a few MHz bandwidth. An obvious solution to circumvent this problem is to lower the emission rate to the kHz regime where single electron tunneling events have been shown to be accessible and measured with high accuracy \cite{Flindt2009,Ubbelohde12}. In that case the WTD is fully accessible but coherent effects are washed out and the physics is dominated by interactions (Coulomb blockade regime). An alternative idea to get information in the time domain is to follow the approach developed by Hong, Ou and Mandel in quantum optics \cite{HOM}. Such an interference experiment between indistinguishable particles created by two identical sources gives access to the second order coherence $g_2(\tau)$ which has been measured in two recent ground breaking experiments \cite{Bocquillon2013,Dubois}. In fact, the WTD and the second order correlation function have the same short time behavior and for a large class of systems (systems with renewal properties which is the case here only for weak overlap) they contain the same amount of information and are related to each other \cite{Cox,Albert2011}. For more general systems, the WTD is a mixture of all the correlation functions but could be reconstructed from their measurement. Unfortunately, only a few of the first ones are measurable with current technology. However, high frequency measurements are progressing very fast \cite{Dubois,Parmentier2012,Basset} and bring some hope to observe electrons one by one in a near future. For instance, the setup described in \cite{Meunier2014} enables us to measure electrons on time scales of fractions of nanoseconds. Although the magnetic field in theses experiments is still too low to reach the single quantum channel limit, they have access to the GHz frequencies needed for WTD measurements.

The last possibility would be to consider alternative systems to electrons where the typical coherence time is much larger. Therefore, coherent effects could be measurable on time scale accessible with present technology. This is the case with cold atoms where the bosonic WTD and FCS have already been measured in atomic lasers \cite{Esslinger}. Fermionic atoms have been a bit more challenging but many progresses have been done in this direction recently \cite{Prentiss1999,Brantut2012,Brantut2014}. In that case there is, to our knowledge, no available protocol to generate Lorentzian pulses yet. However, continuous sources are already available \cite{Brantut2014} and combined with extremely efficient observation tools are very promising setups to investigate WTD. Moreover, such systems present many advantages compared to electronic systems and one of them is the amazing control over interactions and decoherence. 

\section{Conclusion}

We have presented a theory of WTD for periodic trains of quantized pulses impinging on a QPC, with transmission $T_e$. This generalizes WTD for a QPC subjected to a constant voltage \cite{Albert12}. When the pulses weakly overlap, the WTD exhibits strong oscillations on the scale of the period and the decay, at large times, is exponential with a decay rate proportional to $-\ln (1-T_e)$. The internal structure reveals correlations encoded in the many-body state whereas the overall envelope is mainly controlled by the scattering matrix of the QPC. As the overlap between wave packets is increased, the oscillations are reduced to small wiggles. For very large overlaps, the WTD is analogous to that for a QPC subjected to a finite voltage $V$ but with the period $t_p$ playing the role of $h/(eV)$. For $T_e$ not close to $1$, regardless of overlap, the envelope of the WTD is like the one that would be obtained for a binomial process. Electrons cross the QPC randomly with probability $T_e$, every $t_p$. This looks surprising since the quantum system seems to behave like a classical one. For perfect transmission however, quantum correlations are stronger and cause the WTD to show Gaussian decay at large times, reminiscent to what happens for random matrix models. Finally, we have briefly explained how the short time behavior of the WTD could be extracted from Hong-Ou-Mandel experiments. 

Extension of this type of approach to other physical situations such as Klein tunneling, dynamical Coulomb blockade or to the case of several channels could be envisaged. 
 
\acknowledgments

We thank D. Dasenbrook, P. Degiovanni, G. F\`eve, C. Flindt, G. Haack, T. Jonckheere, M. Moskalets and K. H. Thomas for useful discussions and remarks. In addition, we thank C. Flindt for letting us know about a related work \cite{Dasenbrook}.

\appendix

\section{Derivation of Eq. (\ref{generalformula}) of the main text}

In this section, we derive Eq. (\ref{generalformula}) of the main text, giving the idle time probability for arbitrary $N$. Before detecting an electron at $t=0$, the system is in state
\begin{equation}
  \begin{split}
    \psi_s \, = &\, \frac{1}{\sqrt{N!}} 
    \sum_{{\cal P}} (-1)^{\textrm{sgn}({\cal P})}
    \vert \psi_{{\cal P}_{(1)}}(x_1) \rangle    \otimes
    \vert \psi_{{\cal P}_{(2)}}(x_2) \rangle  \\  
    & \otimes
    ... \otimes
    \vert \psi_{{\cal P}_{(N)}}(x_N) \rangle.
\end{split}
\end{equation}
Summation is over all permutations ${\cal P}$ of $(1,2,...,N)$ and
 $\textrm{sgn}({\cal P})$ is the signature of ${\cal P}$.

Applying $Q_1 = \vert x_0 \rangle_1 \langle x_0 \vert_1$,  
\begin{equation}
  \begin{split}
    Q_1 \psi_S \, = & \, 
    \frac{1}{\sqrt{N!}}\, 
    \sum_{{\cal P}} (-1)^{\textrm{sgn}({\cal P})}
        {\tilde \psi}_{{\cal P}_{(1)}}(x_0) \, \vert\varphi_{x_0}(x_1)\rangle \\
        & \otimes 
        \vert \psi_{{\cal P}_{(2)}}(x_2) \rangle
        \otimes
        ... \otimes
        \vert \psi_{{\cal P}_{(N)}}(x_N) \rangle
  \end{split}
\end{equation}

\begin{widetext}
  The wave function after measurement can be recast as a determinant
\begin{equation}
Q \psi_S \, =\, 
\frac{1}{\sqrt{N!}}\, 
\left|
\begin{matrix}
{\tilde \psi}_1(x_0) \varphi_{x_0}(x_1) & 
{\tilde \psi}_2(x_0) \varphi_{x_0}(x_1) & 
... &
{\tilde \psi}_N(x_0) \varphi_{x_0}(x_1) \cr 
\psi_1(x_2) & 
\psi_2(x_2) & 
... & 
\psi_N(x_2) \cr 
... &
... &
... &
... \cr
\psi_1(x_N) & 
\psi_2(x_N) &
... &
\psi_N(x_N)
\end{matrix}
\right|.
\end{equation}
The probability of measuring nothing before $\tau$ is
\begin{equation}
  \begin{split}
    P(\tau,t_0) \, &=\, 
    \langle Q \psi_S \vert 
    \Bigl(1 - \int_{x_0 - v_F \tau}^{x_0} \vert x \rangle \langle x \vert \, dx\Bigr)_2
    \otimes 
    \Bigl(1 - \int_{x_0 - v_F \tau}^{x_0} \vert x \rangle \langle x \vert \, dx\Bigr)_3
    \otimes
    ...\,
    \otimes 
    \Bigl(1 - \int_{x_0 - v_F \tau}^{x_0} \vert x \rangle \langle x \vert \, dx\Bigr)_N
    \vert Q \psi_S \rangle \, \nonumber \\
    &= \,\frac{1}{N!}
    \sum_{{\cal P}} \sum_{{\cal P}^{\prime}} 
    (-1)^{\textrm{sgn}({\cal P})} (-1)^{\textrm{sgn}({\cal P}^{\prime})}
    \prod_{m=1}^N 
    \langle \psi_{{\cal P}_{(m)}}(x_m) \vert Q^{\prime}_m \vert 
    \psi_{{\cal P}^{\prime}_{(m)}}(x_m) \rangle,
  \end{split}
\end{equation}
with $Q^{\prime}_m = (1-Q)$ for $m \not= 1$ and $Q^{\prime}_m = Q(t_u)= \int_{x_0 - v_F t_u}^{x_0} \vert x \rangle \langle x \vert \, dx$, for $m=1$.
For $m=1$, we measure an electron and for $m \not= 1$, we measure no electron.

As usual, the composition of ${\cal P}$ and  ${\cal P}^{\prime}$ has to be considered
 but, contrary to what usually happens, the operator $Q^{\prime}_m$ is not
 the same for all $m$. The case $m=1$ is different from the other $m$'s. We set 
\begin{eqnarray}
{\cal P}^{\prime \prime} \, &=&\, {\cal P} {\rm o} {\cal P}^{\prime}, \\
{\cal P}_1 \, &=& \, {\cal P}^{-1},
\end{eqnarray}
where ``${\rm o}$'' means composition of applications and ${\cal P}^{-1}$ the inverse of ${\cal P}$.
\begin{equation}
  P(\tau,t_0) \, = \,\frac{1}{N!} \, \sum_{{\cal P}_1} \sum_{{\cal P}^{\prime \prime}} (-1)^{\textrm{sgn}({\cal P}^{\prime \prime})}
 \prod_{m=1}^N\langle \psi_m(x_m) \vert Q^{\prime}_{{\cal P}_1(m)} \vert \psi_{{\cal P}^{\prime \prime}(m)}\rangle.
\end{equation}
Among all permutations ${\cal P}_1$ of $(1,2,...,N)$, only $(N-1)!$ will give
${\cal P}_1(m) =1$. 
Finally,
\begin{equation}
P(\tau,t_0) \, =\, 
\frac{1}{N} 
\biggl(
\sum_{k=1}^N 
\textrm{det} \langle \psi_i(x_i) \vert {\tilde Q}_k \vert \psi_j(x_j) \rangle
\biggr),
\label{pitaugeneral}
\end{equation} 
where ${\tilde Q}_k$ means $(1-Q)$ everywhere except for $x_i=k$, where $(1-Q)$ has to be replaced by $Q(t_u)$.         

As an example, here is the term for $k=1$,
\begin{equation}
\textrm{det} \langle \psi_i(x_i) \vert {\tilde Q}_1 \vert \psi_j(x_j) \rangle \, =\, 
\left|
\begin{matrix}
\langle \psi_1 \vert Q(t_u) \vert \psi_1 \rangle &
\langle \psi_1 \vert Q(t_u) \vert \psi_2 \rangle &
... &
\langle \psi_1 \vert Q(t_u) \vert \psi_N \rangle \cr
\langle \psi_2 \vert 1- Q \vert \psi_1 \rangle &
\langle \psi_2 \vert 1-Q  \vert \psi_2 \rangle &
... &
\langle \psi_2 \vert 1-Q \vert \psi_N \rangle \cr
... &
... &
... &
... \cr
\langle \psi_N \vert 1- Q \vert \psi_1 \rangle &
\langle \psi_N \vert 1-Q  \vert \psi_2 \rangle &
... &
\langle \psi_N \vert 1-Q \vert \psi_N \rangle
\end{matrix}
\right|.
\label{termek1}
\end{equation}
The term for $k=2$ is,
\begin{equation}
\textrm{det} \langle \psi_i(x_i) \vert {\tilde Q}_2 \vert \psi_j(x_j) \rangle \, =\, 
\left|
\begin{matrix}
\langle \psi_1 \vert 1-Q \vert \psi_1 \rangle &
\langle \psi_1 \vert 1-Q \vert \psi_2 \rangle &
... &
\langle \psi_1 \vert 1-Q \vert \psi_N \rangle \cr
\langle \psi_2 \vert Q(t_u) \vert \psi_1 \rangle &
\langle \psi_2 \vert Q(t_u)  \vert \psi_2 \rangle &
... &
\langle \psi_2 \vert Q(t_u) \vert \psi_N \rangle \cr
... &
... &
... &
... \cr
\langle \psi_N \vert 1- Q \vert \psi_1 \rangle &
\langle \psi_N \vert 1-Q  \vert \psi_2 \rangle &
... &
\langle \psi_N \vert 1-Q \vert \psi_N \rangle
\end{matrix}
\right|.
\label{termek2}
\end{equation}
Normalization of $\Pi(\tau,t_0)$ requires that one has to divide by the same sum 
 of determinants, but now, the $1-Q$ have to be replaced by $1$.  
This gives Eq. (\ref{generalformula}) of the main text.

\section{Weak overlap and quasi-diagonal matrices}

In this section, we justify why, in the case of small overlap, the approximation by diagonal matrices works so well.
If we forget the logarithmic terms, for 
$3 t_p <  \tau <  4 t_p$, $R_{\ell,\ell^{\prime}} - T_e N_{\ell,\ell^{\prime}}$ becomes
\begin{equation}
\begin{pmatrix}
\rho_1      & \rho_1 \frac{2 \pi R }{ 2 \pi R -i}    &  \rho_1 \frac{2 \pi R }{ 2 \pi R - 2i} & 
\rho_2 \frac{2 \pi R }{ 2 \pi R - 3i}    &  \rho_2 \frac{2 \pi R }{ 2 \pi R - 4i}  &  ...  &  
 \rho_2 \frac{2 \pi R }{ 2 \pi R - \ell i}
 & ...    \cr  
\rho_1 \frac{2 \pi R }{ 2 \pi R +i}  &  \rho_1    & \rho_1 \frac{2 \pi R }{ 2 \pi R -i} & 
\rho_2 \frac{2 \pi R }{ 2 \pi R -2i}
  & 
\rho_2\frac{2 \pi R }{ 2 \pi R -3i}
  &  ...  & 
\rho_2 \frac{2 \pi R }{ 2 \pi R -\ell i}
  & .... \cr
\rho_1 \frac{2 \pi R }{ 2 \pi R + 2i} &   \rho_1 \frac{2 \pi R }{ 2 \pi R + i}  & \rho_1  &
 \rho_2 \frac{2 \pi R }{ 2 \pi R - i}
  & 
\rho_2 \frac{2 \pi R }{ 2 \pi R - 2i}
&  ...  & 
\rho_2 \frac{2 \pi R }{ 2 \pi R - \ell i}
 & ... \cr
\rho_2 \frac{2 \pi R }{ 2 \pi R + 3i} &  
\rho_2 \frac{2 \pi R }{ 2 \pi R + 2i} & 
\rho_2 \frac{2 \pi R }{ 2 \pi R + i} & 
 1   &  \frac{2 \pi R }{ 2 \pi R - i}   &  \frac{2 \pi R }{ 2 \pi R -2i} 
 & \frac{2 \pi R }{ 2 \pi R -3i}    & ... \cr
\rho_2 \frac{2 \pi R }{ 2 \pi R + 4i}  & 
\rho_2 \frac{2 \pi R }{ 2 \pi R + 3i}  & 
\rho_2 \frac{2 \pi R }{ 2 \pi R + 2i}  & 
\frac{2 \pi R }{ 2 \pi R + i}   & 1   &  \frac{2 \pi R }{ 2 \pi R -i} & 
\frac{2 \pi R }{ 2 \pi R - 2 i}   & ... \cr
...      &  ...  &  ...  &
 \frac{2 \pi R }{ 2 \pi R +2i}   &  \frac{2 \pi R }{ 2 \pi R +i}  &  1  & 
\frac{2 \pi R }{ 2 \pi R -i} & ...\cr
\rho_2\frac{2 \pi R }{ 2 \pi R + \ell^{\prime} i}     & 
\rho_2\frac{2 \pi R }{ 2 \pi R + (\ell^{\prime}-1) i} &  
\rho_2\frac{2 \pi R }{ 2 \pi R + (\ell^{\prime}-2) i} & 
\frac{2 \pi R }{ 2 \pi R + (\ell^{\prime} - 3)i}   & ...   &  \frac{2 \pi R }{ 2 \pi R + i} &
\delta_{\ell,\ell^{\prime}}   & ... \cr
 \end{pmatrix}
\end{equation}
with $\rho_1=1-T_e$ and $\rho_2=1-T_e/2$. Thus, in the limit $R \rightarrow 0$, 
$\textrm{det} (R_{\ell,\ell^{\prime}} - T_e N_{\ell,\ell^{\prime}})$ factorizes as $(1-T_e)^{\textrm{int}(\tau/t_p)}
 \textrm{det}({\tilde R}_{\ell,\ell^{\prime}})$, 
where ${\tilde R}_{\ell,\ell^{\prime}}$ 
is the same as 
$R_{\ell,\ell^{\prime}}$
 except that all elements 
 with row {\it or} column index smaller than $\textrm{int}(\tau/t_p)$ have to be replaced by the elements 
of the identity matrix. 

\end{widetext}

So far, we have neglected the logarithmic terms. They vanish on the diagonal. They also vanish for $\ell$ or $\ell^{\prime}$ much smaller than $\tau/t_p$ and also when $\ell$ and $\ell^{\prime}$ much larger than $\tau/t_p$. Neglecting them should amount to neglecting some correlations and the decay rate including them may be smaller than the one given in the diagonal approximation, but we do not have clear evidence for that. Calculations of the first non-zero order in $R$ would require retaining all the terms in Eq. (\ref{generalformula}) of the main text.


\begin{thebibliography}{99}
\bibitem{notex0} In this limit, one can easily show that $x_0$ and $t_0$ scale out of the problem.
\end{thebibliography}
\end{document}